\documentstyle[aps,multicol,epsfig]{revtex}
\tighten

\begin{document}
\title{Quantum distribution functions for radial observables}
\author{J. Twamley}
\address{Theory and Laser Optics Groups\\
Blackett Lab, Imperial College,\\
London, SW7 2BZ, UK}
\date{8 December 1996}
\maketitle

\begin{abstract}
For quantum systems with two dimensional configuration space we construct a
physical radial momentum observable. Rescaling the radius we find the
dilatonic degrees of freedom form a Weyl algebra. With this we construct the
radial Wigner quasi-probability distribution function.
\end{abstract}

\begin{multicols}{2}
The Wigner quasi-probability distribution function is a familiar tool to
many working in quantum and atom optics \cite{HILLERY}. It is primarily used
in the classical-quantum correspondence where the appearance of positive and
negative regions of the Wigner function gives easily understood information
concerning the probability concentrations and quantum interferences present
within the quantum state \cite{SCHLEICH}. Typically one describes the Wigner
function on a phase space which is labelled by cartesian coordinates of
position and momentum. For physical systems which admit a two dimensional
cylindrical symmetry, eg. trapped ultra-cold ions, bose condensates, etc.,
clearly a polar description of the Wigner function would be more natural.
However, no such description has appeared in the literature. In this letter
we show, in three stages, how a Wigner function for the radial observables
can be constructed. This radial Wigner function could be reconstructed from
experimental data much as recent experiments have reconstructed the
cartesian Wigner function for the one dimensional motion of a trapped ion 
\cite{WINELAND}. The angular parts of the complete four dimensional Wigner
function are complicated by the imposition of singlevaluedness of the
wavefunction under a rotation of $2\pi $ which cause the conjugate angular
momentum to become discrete. We leave the angular part for a later work.

The stages towards the construction of a radial-Wigner function proceed as
follows: (1) a proper Wigner function possesses marginals which are true
probability distributions for the observables whose eigenvalues label the
Wigner function and thus the phase space axes. For a single degree of
freedom a mere transformation of the cartesian position and momentum into
polar form does not yield a proper Wigner function for the polar observables 
\cite{KNIGHT}. This is also true for higher dimensional phase space
representations. Central to the problem is the correct specification of the
radial momentum operator. By noting the symmetry action of the momentum on
the half-infinite radius observable we construct a physical ``conjugate''
momentum $\hat{P}^{r}$; (2) essential to the construction of the Wigner
function is the existence of ``point'' operators $\hat{A}(\lambda
_{1},\lambda _{2})$, which obey, ${\rm Tr}[\hat{A}(\lambda _{1},\lambda _{2})%
\hat{A}^{\dagger }(\lambda _{1}^{\prime },\lambda _{2}^{\prime })]=\delta
(\lambda _{1}-\lambda _{1}^{\prime })\delta (\lambda _{2}-\lambda
_{2}^{\prime })$ \cite{WOOTTERS,CAHILL}. We find that the radial $\hat{A}%
(r,P^{r})$, is not completely exponential in the radial position and
momentum operators; (3) guided by the form of $\hat{A}(r,P^{r})$ we make the
coordinate transformation $\hat{v}\equiv \ln \hat{r}$. We find, $[\hat{v},%
\hat{P}^{r}]=i\hbar $, the Weyl algebra. After re-scaling the eigen-basis
kets of $\hat{v}$ to recover the standard resolution of unity we can easily
construct the radial Wigner function, $W(v,P^{r})$. This radial (or
dilation) Wigner function gives the proper marginal probability
distributions for $\hat{v}$ and $\hat{P}^{r}$. We also note the existence of
a dilaton ground state $|0\rangle _{dilations}$, and calculate the
wavefunction of this ground state in the $\hat{r}$ basis. We finally
calculate the radial Wigner function for the lowest Schwinger states $%
|l,0\rangle $, (these are simultaneous eigenstates of energy and angular
momentum), and briefly outline how, given a quantum state in a two
dimensional harmonic Fock representation, one can construct the radial
reduced density matrix, and from there the radial-Wigner function.

Dirac in his textbook on quantum mechanics \cite{DIRAC} introduced the
following momentum, conjugate to the radial coordinate 
\begin{equation}
\hat{P}^{D}\equiv \frac{1}{\hat{r}}\left( \hat{x}\hat{P}_{x}+\hat{y}\hat{P}%
_{y}-i\hbar /2\right) \;\;.  \label{DIRAC2}
\end{equation}
The factor of two difference in the $\hbar $ term arrises because we are
working in two dimensions instead of the three in Dirac's case. Using the
cartesian commutation relations between $\hat{x}$ and $\hat{P}_{x}$ one can
easily confirm that 
\begin{equation}
\lbrack \hat{r},\hat{P}^{D}]=i\hbar \;\;.  \label{diraccom}
\end{equation}
Using the Baker-Campbell-Hausdorff (CBH) expansion and (\ref{diraccom}) we
can show 
\begin{equation}
\exp (i\varsigma \hat{P}^{D}/\hbar )\;\hat{r}\;\exp (-i\varsigma \hat{P}%
^{D}/\hbar )=\hat{r}+\varsigma \;\;.  \label{eq1}
\end{equation}
Since $\hat{r}$ has a half-infinite spectrum, the operator $\hat{P}^{D}$
cannot be self-adjoint. Although this has been noted by many others \cite
{MESSIAH}, no self-adjoint radial momentum has been proposed that the author
can find. Before going on to construct a self-adjoint radial momentum we
first calculate the representation of $\hat{P}^{D}$ in the $\hat{r}$
eigen-ket basis (i.e. $\hat{r}|r\rangle =r|r\rangle $). From the commutation
relation (\ref{diraccom}) we must have $\langle r|\hat{P}^{D}|\psi \rangle
\sim -i\hbar (\partial _{r}+f(r))\langle r|\psi \rangle $. From (\ref{DIRAC2}%
), and expressing the cartesian partial derivatives in polar coordinates we
find 
\begin{equation}
\langle r|\hat{P}^{D}|\psi \rangle =-i\hbar \left( \partial _{r}+\frac{1}{2r}%
\right) \langle r|\psi \rangle \;\;.  \label{represent}
\end{equation}
To obtain a self-adjoint momentum operator we must find an operator which
respects the half-infinite spectrum of $\hat{r}$. This can be achieved if we
have $[\hat{r},\hat{P}^{r}]=i\hbar \hat{r}$, the Sack dilation algebra \cite
{STACK}. Using the CBH expansion we can easily show 
\begin{equation}
\exp (i\varsigma \hat{P}_{r}/\hbar )\;\hat{r}\;\exp (-i\varsigma \hat{P}%
_{r}/\hbar )=\hat{r}e^{\varsigma }\;\;.  \label{expcbh}
\end{equation}
To explicitly construct a $\hat{P}^{r}$ we form 
\begin{equation}
\hat{P}^{r}\equiv \hat{r}\hat{P}^{D}-\frac{i\hbar }{2}\;,  \label{rad-dirac}
\end{equation}
where the additional $\hbar /2$ makes $\hat{P}^{r}$ hermitian. To obtain the
representation of $\hat{P}^{r}$ in the $|r\rangle $ basis we simply use (\ref
{represent}) above to find 
\[
\langle r|\hat{P}^{r}|\psi \rangle =-i\hbar \left( r\partial _{r}+1\right)
\langle r|\psi \rangle \;\;. 
\]
From this we can calculate the transition function $\langle r|P^{r}\rangle $
where $|P^{r}\rangle $ is the eigen-ket of the operator $\hat{P}^{r}$ to be 
\begin{equation}
\langle r|P^{r}\rangle =\frac{1}{\sqrt{2\pi \hbar }}\frac{r^{iP^{r}/\hbar }}{%
r}\;\;.
\end{equation}
This is normalised with the measure $rdr$ and gives 
\[
\smallint _{0}^{\infty }rdr\;\langle r|P^{r}\rangle \langle P^{\prime
r}|r\rangle =\delta \left( P^{r}-P^{\prime r}\right) \;\;. 
\]
The action of $\hat{r}$ on $\hat{P}^{r}$ is not a simple scaling, instead 
\[
e^{-i\varsigma \hat{r}/\hbar }\hat{P}^{r}e^{i\varsigma \hat{r}/\hbar }=\hat{P%
}^{r}+\varsigma \hat{r}\;\;. 
\]
To form a displacement, or point, operator we might be tempted to
exponentiate a linear combination of the position and momentum operators in
analogy with the harmonic oscillator. However this construction does not
result in a unique adjoint action of the displacement operator on $\hat{r}$
and $\hat{P}^{r}$. From \cite{WITSCHEL} we see 
\begin{eqnarray}
{\cal D}(a,m) &\equiv &\exp \left( \frac{ia}{\hbar }\left[ \hat{P}^{r}+m\hat{%
r}\right] \right) \\
&=&\exp \left( \frac{ia}{\hbar }\hat{P}^{r}\right) \exp \left[ \frac{im\hat{r%
}}{\hbar }\left( 1-e^{-a}\right) \right]  \nonumber \\
&=&\exp \left[ \frac{im\hat{r}}{\hbar }\left( e^{a}-1\right) \right] \exp
\left( \frac{ia}{\hbar }\hat{P}^{r}\right) \;\;.  \nonumber
\end{eqnarray}
Thus the choice of ordering alters the action of ${\cal D}(a,m)$ on the
operators $\hat{r}$ and $\hat{P}^{r}$. This is to be expected since the
algebra is not symmetric in $\hat{r}$ and $\hat{P}^{r}$. One must explicitly
specify the ordering in this displacement operator as is done in most
treatments of displacement operators for spin systems.

An essential property for a displacement operator is \cite{WOOTTERS,CAHILL}, 
\begin{equation}
Tr[D(\lambda ,\mu )D^{\dagger }(\lambda ^{\prime },\mu ^{\prime })]\approx
\delta (\lambda -\lambda ^{\prime })\delta (\mu -\mu ^{\prime })\;\;.
\label{Dis-delta}
\end{equation}
No operator in the form of products of exponentials of $\hat{r}$ and $\hat{P}%
^{r}$ will satisfy (\ref{Dis-delta}). We instead consider the operator \cite
{LOUISELL}, 
\begin{equation}
D(\lambda ,\mu )\equiv \exp (i\mu \hat{P}^{r}/2\hbar)\;r^{i\lambda }\exp
(i\mu \hat{P}^{r}/2\hbar)\;\;,  \label{exp-displace}
\end{equation}
where $\lambda \in [0,+\infty )$. Taking the trace over $|r\rangle $ with
measure $\frac{rdr}{2\pi },$ from $r=0$ to $r=+\infty $ gives 
\begin{equation}
Tr[D(\lambda ,\mu )D^{\dagger }(\lambda ^{\prime },\mu ^{\prime })]=\frac{1}{%
e^{2\mu }}\delta (\lambda -\lambda ^{\prime })\delta (\mu -\mu ^{\prime
})\;\;.  \label{delta}
\end{equation}
One can show $D^{\dagger }(\lambda ,\mu )\hat{r}D(\lambda ,\mu )=\hat{r}%
e^{-\mu }$. However to evaluate $D^{\dagger }(\lambda ,\mu )\hat{P}%
^{r}D(\lambda ,\mu )$ one must calculate $\hat{r}^{\alpha }\hat{P}^{r}\hat{r}%
^{-\alpha }$. This is done by expressing $\hat{P}^{r}$ in terms of $\hat{P}%
^{D}$ through (\ref{rad-dirac}), using the simpler commutation relations for 
$\hat{P}^{D}$, the rule $[f(\hat{r}),\hat{P}^{D}]=i\hbar \partial f(\hat{r}%
)/\partial \hat{r}$, and finally transforming back to $\hat{P}^{r}$. This
surprisingly gives $D^{\dagger }(\lambda ,\mu )\hat{P}^{r}D(\lambda ,\mu )=%
\hat{P}^{r}+\lambda $.

Thus, by using the new ``displacement'' operator (\ref{exp-displace}), one
can (except for the exponential factor in (\ref{delta})), almost completely
regain the properties of the Harmonic oscillator displacement operator.
Although one could now proceed to construct a Wigner function with this
displacement operator, the replacement of $\exp (i\lambda \hat{r})$ with $%
\hat{r}^{i\lambda }$ in (\ref{exp-displace}), and the exponential prefactor
in (\ref{delta}) point towards a clearer understanding of the situation.
Guided by these signals we now make the transformation to a new coordinate
operator, 
\begin{equation}
\hat{v}\equiv \ln \hat{r}\;\;.
\end{equation}
Using the definition of $\hat{P}^{r}$ in terms of $\hat{P}^{D}$ (\ref
{rad-dirac}), we can show 
\begin{equation}
\lbrack \hat{v},\hat{P}^{r}]=[\ln \hat{r},\hat{P}^{r}]=[\ln \hat{r},\hat{r}%
\hat{P}^{D}]=\hat{r}[\ln \hat{r},\hat{P}^{D}]=\hat{r}\frac{i\hbar }{\hat{r}}%
=i\hbar \;\;.  \label{v-comm}
\end{equation}
With this we have re-gained the algebra of the original cartesian position
and momentum observables. Since the spectrum for both $\hat{v}$ and $\hat{P}%
^{r}$ ranges from $(-\infty ,+\infty )$ the translation generating Weyl
algebra in (\ref{v-comm}) properly respects the spectrum of both operators
with the result that both $\hat{v}$ and $\hat{P}^{r}$ are self-adjoint and
represent physical observables. We can construct creation and annihilation
operators in the same manner as the harmonic oscillator, $\hat{a}\equiv (%
\hat{v}+i\hat{P}^{r})/\sqrt{2}$, and $\hat{a}^{\dagger }\equiv (\hat{v}-i%
\hat{P}^{r})/\sqrt{2}$. In particular, there exists a proper vacuum state $%
|0\rangle _{dilations}$ which is annihilated by $\hat{a}$.

Not everything is straightforward however. The resolution of unity in the $%
|r\rangle $ basis becomes, in the eigenbasis of $\hat{v}$, 
\begin{eqnarray}
2\pi 1\!\!1&=&\int_{-\infty }^{+\infty }dx\;|x\rangle \langle
x|=\int_{0}^{\infty }rdr\;|r\rangle \langle r| \\
&=&\int_{-\infty }^{+\infty }e^{2v}dv\;|v\rangle \langle v|\;\;.
\end{eqnarray}
With this resolution of unity we must have 
\begin{equation}
1\!\!1|v^{\prime }\rangle =|v^{\prime }\rangle =\frac{1}{2\pi}\int_{-\infty
}^{+\infty }e^{2v}dv\;|v\rangle \langle v|v^{\prime }\rangle \;\;,
\end{equation}
and thus $\langle v|v^{\prime }\rangle \equiv \frac{1}{e^{2v}}\delta
(v-v^{\prime })=2\delta (e^{2v}-e^{2v^{\prime }})$. This exponential measure
complicates matters and so to simplify we rescale the basis kets to be 
\begin{equation}
|\bar{v}\rangle ^{\bullet }\equiv e^{\bar{v}}|e^{\bar{v}}\rangle _{r}
\label{vbar-def}
\end{equation}
where $|e^{\bar{v}}\rangle _{r}$ is a eigenket of $\hat{r}$ with eigenvalue $%
e^{\bar{v}}$. The resolution of the identity in the new $|\bar{v}\rangle
^{\bullet }$ basis is, $2\pi 1\!\!1=\int_{-\infty }^{+\infty }d\bar{v}\;|%
\bar{v}\rangle ^{\bullet }{}^{\bullet }\langle \bar{v}|$. The inner product
between $|\bar{v}\rangle ^{\bullet }$ basis kets is now simply ${}^{\bullet
}\langle \bar{v}|\bar{v}^{\prime }\rangle ^{\bullet }\equiv \delta (\bar{v}-%
\bar{v}^{\prime })$. This rescaling will primarily appear in the calculation
of ${}^{\bullet }\langle \bar{v}|\psi \rangle $ when we have $\langle r|\psi
\rangle $. To be certain that everything is identical to the Harmonic
oscillator case we finally evaluate ${}^{\bullet }\langle \bar{v}|\hat{P}%
^{r}|\psi \rangle $. Since $|\bar{v}\rangle ^{\bullet }=r|r\rangle $ where $%
r=e^{\bar{v}}$, we have 
\begin{eqnarray}
{}^{\bullet }\langle \bar{v}|&\hat{P}^{r}&|\psi \rangle =r\langle r|\hat{P}%
^{r}|\psi \rangle |_{r=e^{\bar{v}}} =\left. r\left[ -i\hbar \left( r\partial
_{r}+1\right) \right] \langle r|\psi \rangle \right| _{r=e^{\bar{v}}} \\
&=&\left. r\left[ -i\hbar \left( r\partial _{r}+1\right) \frac{1}{r}\right]
{}^{\bullet }\langle \bar{v}|\psi \rangle \right| _{r=e^{\bar{v}}}  \nonumber
\\
&=&\left. -i\hbar r\partial _{r}{}^{\bullet }\langle \bar{v}|\psi \rangle
\right| _{r=e^{\bar{v}}}=-i\hbar \partial _{\bar{v}}{}^{\bullet }\langle 
\bar{v}|\psi \rangle \;\;.  \nonumber
\end{eqnarray}
We see that the momentum operator $\hat{P}^{r}$, in the basis $|\bar{v}%
\rangle ^{\bullet }$ is the familiar $-i\hbar \partial _{\bar{v}}$, the same
as the cartesian case.

As in the Harmonic oscillator case we define the Wigner and s-ordered
quasi-distribution function to be 
\begin{equation}
W(\xi ,s)\equiv \int \frac{d^{2}\alpha }{\pi }\;e^{\alpha \xi ^{*}-\alpha
^{*}\xi }\;C(\alpha ,s)\;\;,
\end{equation}
where $C(\alpha ,s)$ is the s-ordered displacement operator 
\begin{equation}
C(\alpha ,s)\equiv Tr\left[ \rho D(\alpha )e^{s|\alpha |^{2}/2}\right] \;\;,
\end{equation}
and $D(\alpha )\equiv \exp (\alpha \hat{a}^{\dagger }-\alpha ^{*}\hat{a})$,
where $\hat{a}^{\dagger }$ and $\hat{a}$ , the dilatonic creation and
annihilation operators, are defined as above. For $s=0$, $\xi =\gamma
+i\delta $, the Wigner function becomes 
\begin{equation}
W(\gamma ,\delta )=\frac{1}{2\pi }\int d\varepsilon \;e^{-i\varepsilon
\delta }\;{}^{\bullet }\langle \gamma +\frac{\varepsilon }{2}|\rho |\gamma -%
\frac{\varepsilon }{2}\rangle ^{\bullet }\;\;.  \label{wig1}
\end{equation}
We now calculate the Wigner function for the quantum states $\rho _{l}\equiv
|l,0\rangle \langle l,0|$, where $|l,m\rangle $ is the Schwinger angular
momentum state given by 
\begin{equation}
|l,m\rangle \equiv \frac{1}{\sqrt{(l+m)!(l-m)!}}\hat{A}_{+}^{\dagger (l+m)}%
\hat{A}_{-}^{\dagger (l-m)}|0,0\rangle \;\;,
\end{equation}
where $\hat{A}_{+}\equiv (\hat{a}_{x}-i\hat{a}_{y})/\sqrt{2}$, $\hat{A}%
_{-}\equiv (\hat{a}_{x}+i\hat{a}_{y})/\sqrt{2}$ \cite{TANNOUDJI}. Either by
solving the generating differential equation (as is done in \cite{TANNOUDJI}%
) or by solving the harmonic oscillator radial equation one finds $\langle
r,\phi |l,m\rangle =R_{l,m}(r)\Theta _{m}(\phi )$, where 
\end{multicols}
\noindent\rule{0.5\textwidth}{0.4pt}\rule{0.4pt}{\baselineskip}
\begin{eqnarray}
R_{l,m}(r) &=&\beta \sqrt{\frac{2(l-|m|)!}{(l+|m|)!}}(\beta
r)^{2|m|}\;e^{-\beta ^{2}r^{2}/2}\;{\rm L}_{l-|m|}^{2|m|}(\beta
^{2}r^{2})\;(-1)^{l-|m|}\;\;, \\
\Theta _{m}(\phi ) &=&\frac{1}{\sqrt{2\pi }}\;e^{2im\phi }\;\;,
\end{eqnarray}
$L_{n}^{\alpha }$ is the associated Laguerre polynomial and $\beta =1/\sqrt{%
\hbar }$. Integrating out the $\phi $ dependence, setting $\hbar =1$ and
using (\ref{vbar-def}) we obtain 
\begin{equation}
{}^{\bullet }\langle \bar{v}|l,0\rangle =\sqrt{2}e^{\bar{v}}e^{-e^{2\bar{v}%
}/2}{\rm L}_{l}^{0}(e^{2\bar{v}})(-1)^{l}\;\;.  \label{vbarprod}
\end{equation}
Inserting (\ref{vbarprod}) into (\ref{wig1}) we finally get 
\begin{equation}
W_{l}(\gamma ,\delta )=\frac{2e^{2\gamma }}{\pi }\int_{-\infty }^{+\infty
}d\varepsilon \;e^{-2i\varepsilon \delta }\exp \left[ -e^{2\gamma }\cosh
2\varepsilon \right] \;{\rm L}_{l}(e^{2(\gamma +\varepsilon )}){\rm L}%
_{l}(e^{2(\gamma -\varepsilon )})\;\;.  \label{finalwig}
\end{equation}
\hspace*{\fill}\rule[0.4pt]{0.4pt}{\baselineskip}%
\rule[\baselineskip]{0.5\textwidth}{0.4pt}  
\begin{multicols}{2}%
The radial
Wigner function, (\ref{finalwig}), is numerically integrated for the first
four $|l,0\rangle $ states and plotted in Figs. 1(a)-1(d). The presence of
negative regions in these Wigner functions is not surprising as the pure
states $|l,0\rangle $ are reminiscent of Fock states. One can construct the
dilatonic coherent states $|\alpha \rangle \equiv D(\alpha )|0\rangle
_{dilatons}$. More interesting is $\langle r|0\rangle _{dilatons}$. From the
definition $\hat{a}|0\rangle _{dilatons}=0$, and $\sqrt{2}\hat{a}=\hat{v}+i%
\hat{P}^{r}$, we can evaluate ${}^{\bullet }\langle \bar{v}|\hat{a}|0\rangle
_{dilatons}=0$ to get ${}^{\bullet }\langle \bar{v}|0\rangle
_{dilatons}=N\exp (-\bar{v}^{2}/2)$, where $N^{2}=\sqrt{\pi }$. Again using (%
\ref{vbar-def}) we obtain 
\begin{equation}
\langle r|0\rangle _{dilatons}=\frac{1}{r\sqrt{\pi }}\exp (-\frac{1}{2}\ln
r\ln r)=\frac{r^{-\frac{1}{2}\ln r}}{r\sqrt{\pi }}\;\;,  \label{weird}
\end{equation}
which is normalised to unity, $\int_{0}^{+\infty }rdr\;|\langle r|0\rangle
_{dilatons}|^{2}=1$.

In this letter we examined the problem of constructing a Wigner function for
the two dimensional radial subspace of a quantum system possessing two
continuous degrees of freedom (or a four dimensional Wigner function). Since
the radial subspace is labelled by an operator with an half-infinite
spectrum, previous attempts to define a physical conjugate momentum have
failed. By choosing a momentum operator which respected the spectrum of the
radial coordinate we found we could construct a physically meaningful radial
conjugate momentum. With a logarithmic relabelling of the radial coordinate
the new coordinate and momentum satisfied a Weyl algebra and we were thus
able to carry over all the techniques associated with the harmonic
oscillator to construct a Wigner function. One can identify a ground state
for these observables which is destroyed by a dilatonic annihilation
operator. We finally examined the radial-Wigner function for particular
quantum states and found the radial wavefunction for the dilatonic vacuum
state. To examine the radial-Wigner function for more general quantum states
is difficult. If one has the density matrix in a $x-y$ Fock state basis, one
must first transform $\rho $ into the Schwinger basis of angular momentum
kets, $\rho =\sum C_{lm;l^{\prime }m^{\prime }}|l,m\rangle \langle l^{\prime
},m^{\prime }|$. One can then find $\langle r,\phi |\rho |r^{\prime },\phi
^{\prime }\rangle $. Finally one must trace over $\phi $. However, even
though we have traced out over $\phi $ one will still end up with a sum over 
$m$. This is so because, in contrast with the cartesian Fock state
decomposition, the radial wavefunction is labelled by {\em both} quantum
numbers $l$ and $m$ while the angular part is only labelled by $m$. We also
note that the Wigner function here gives the correct marginals for the
operators $\hat{P}^{r}$ and $\hat{v}$. Once these marginals and their
corresponding integration measures are obtained one can rescale $v$ to $r$.
One cannot rescale the axis in Fig. 1 to find the Wigner function for $\hat{P%
}^{r}$ and $\hat{r}$ without altering the integration measure and thus, the
function form of the Wigner function. Finally, although an observable, the
radial operator $\hat{v}=\ln \hat{r}$, may not be easy to measure directly.
However, its should be possible, through standard reconstruction techniques,
to numerically approximate the radial Wigner function from experimental data.

The author thanks V. Bu\v {z}ek of Imperial College, London, UK and A. Hurst
at the University of Adelaide, South Australia, for enlightening
discussions. This work was supported through a European Human Capital and
Mobility Fellowship.

\begin{minipage}{3.375in}

\begin{figure}
\begin{center}
\setlength{\unitlength}{1.cm}
\begin{picture}(6,12)
\put(-.2,0.1){\epsfig{file=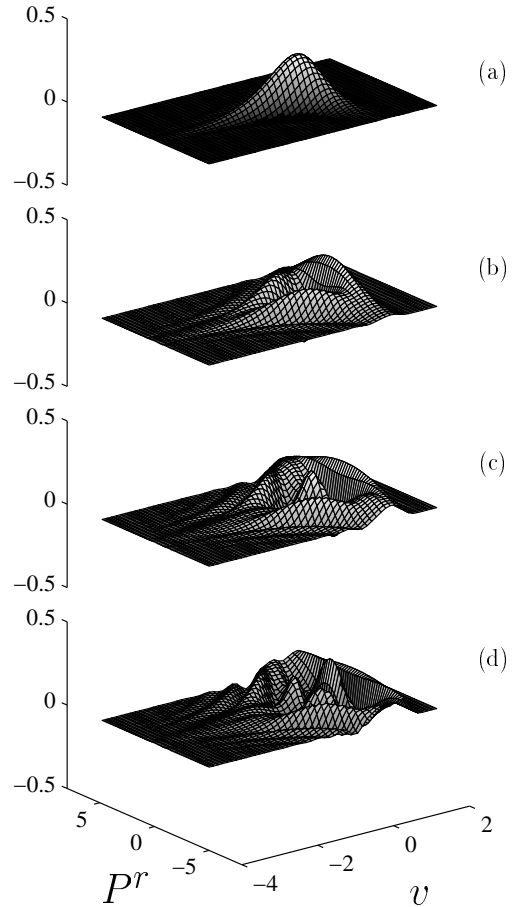,width=7cm}}
\end{picture}
\end{center}
\caption{(a)-(d) Radial Wigner Functions for the states $|l,0\rangle$, 
$l=0,..,3$.}

\end{figure}

\end{minipage}

\end{multicols}

\end{document}